\documentclass[]{spie}  %>>> use for US letter paper
\usepackage{amsmath,amsfonts,amssymb}
\usepackage{graphicx}
\usepackage{subcaption}
\usepackage{multirow}
\usepackage{float}
\usepackage{placeins}
\usepackage[colorlinks=true, allcolors=blue]{hyperref}
\usepackage{nowidow}
\title{A parameter scan of dark zone maintenance for high-contrast imaging of exoplanets using theoretical and experimental implementations}

\author[a]{Saikrishna Manojkumar}
\author[a]{Christine L. Page}
\author[b]{Leonid Pogorelyuk}
\author[c,d]{Susan F. Redmond}
\author[a]{Ajay S. Gill}
\author[e]{Laurent Pueyo}
\author[e]{Emiel H. Por}
\author[f]{Iva Laginja}
\author[e]{Rapha\"el Pourcelot}
\author[e]{Bryony F. Nickson}
\author[e]{Ananya Sahoo}
\author[e]{Meiji M. Nguyen}
\author[e]{R\'emi Soummer}
\author[e]{Marshall D. Perrin}
\author[g]{Bijan Nemati}
\author[a]{Kerri Cahoy}
\author[h]{Jeremy N. Kasdin}

\affil[a]{Massachusetts Institute of Technology, 77 Massachusetts Ave, Cambridge, MA 02139, USA}
\affil[b]{Rensselaer Polytechnic Institute, 110 8th St, Troy, NY 12180, USA}
\affil[c]{California Institute of Technology, 1216 E California Blvd, Pasadena, CA 91125, USA}
\affil[d]{Jet Propulsion Laboratory, 4800 Oak Grove Dr, Pasadena, CA 91011, USA}
\affil[e]{Space Telescope Science Institute, 3700 San Martin Drive, Baltimore, MD 21218, USA}
\affil[f]{LESIA, Observatoire de Paris, Université PSL, Sorbonne Université, Université Paris Cité, CNRS, 5 pl. Jules Janssen, 92195 Meudon, France}
\affil[g]{The University of Alabama in Huntsville, 301 Sparkman Dr NW, Huntsville, AL 35899, USA}
\affil[h]{Princeton University, Olden Street, Princeton, NJ 08544, USA}

\authorinfo{Further author information: (Send correspondence to S.M.)\\S.M.: E-mail: saimanoj@mit.edu}

\begin{document} 
\maketitle
\begin{abstract}

Maintaining wavefront stability while directly imaging exoplanets over long exposure times is an ongoing problem in the field of high-contrast imaging. Robust and efficient high-order wavefront sensing and control systems are required for maintaining wavefront stability to counteract mechanical and thermal instabilities. Dark zone maintenance (DZM) has been proposed to address quasi-static optical aberrations and maintain high levels of contrast for coronagraphic space telescopes. To further experimentally test this approach for future missions, such as the Habitable Worlds Observatory, this paper quantifies the differences between the theoretical closed-loop contrast bounds and DZM performance on the High-contrast Imager for Complex Aperture Telescopes (HiCAT) testbed. The quantification of DZM is achieved by traversing important parameters of the system, specifically the total direct photon rate entering the aperture of the instrument, ranging from $1.85 \times 10^{6}$ to $1.85 \times 10^{8}$ photons per second, and the wavefront error drift rate, ranging from $\sigma_{drift} = 0.3-3~nm/\sqrt{iteration}$, injected via the deformable mirror actuators. This is tested on the HiCAT testbed by injecting random walk drifts using two Boston Micromachines kilo deformable mirrors (DMs). The parameter scan is run on the HiCAT simulator and the HiCAT testbed where the corresponding results are compared to the model-based theoretical contrast bounds to analyze discrepancies. The results indicate an approximate one and a half order of magnitude difference between the theoretical bounds and testbed results.
\end{abstract}


\section{INTRODUCTION}
\label{sec:intro}  % \label{} allows reference to this section

\subsection{Direct Imaging with Space Telescopes}
\label{subsec:1.1}

Current recommendations from the Decadal Survey on Astronomy and Astrophysics 2020 (Astro2020) advocate for the use of large aperture space telescopes to directly image exoplanets of interest while searching for life signatures. This has led to the proposal of the Habitable Worlds Observatory (HWO), to view candidate planets. Viewing these exoplanets would require the instrument to block out light from the host star which can be $10^{10}$ times brighter than the planet light \cite{stark_exoearth_2019}. Starlight suppression of $10^{-10}$ has been demonstrated by NASA using a coronagraph, which is usually composed of one mask in the focal plane to occult starlight and one in the pupil plane to attenuate diffraction effects, combined with two deformable mirrors (DMs)\cite{seo_testbed_2019}. This configuration allows for the development of a high contrast region in the image, known as a dark zone (DZ), by overcoming static optical aberrations present in the system. However, directly imaging exoplanets using a space telescope over long exposure times requires the implementation of high-order wavefront sensing and control (HOWFSC) systems to counteract mechanical and thermal instabilities. 

The upcoming Nancy Grace Roman Space Telescope (RST) will utilize a coronagraph to achieve its maximum contrast of about $10^{-8}$ by pointing at a bright star as a reference to generate a DZ and then slewing over to the dimmer target star to observe. During this observation of the target star, the contrast in the DZ will degrade due to quasi-static aberrations, and the telescope will periodically slew back to the reference star to get DM commands to null these new aberrations\cite{kasdin_nancy_2020}. This approach reduces the observing time on the target star and may not maintain a contrast of $10^{-10}$ for the desired duration of time on large aperture telescopes. The Decadal Survey Testbed (DST) at NASA Jet Propulsion Laboratory (JPL) showed a DZ on the order of $10^{-10}$ deteriorating within an hour when the DM shape was held constant.\cite{meeker_twin_2021} Additionally, repeated slewing may introduce optical instabilities that cannot be easily sensed. 

HOWFSC systems have been shown in theory and experiments to maintain high contrast in the science image through the use of the science exposures as measurements that drive the control and change the shape of DMs on the telescope in a recursive algorithm known as dark zone maintenance (DZM)\cite{pogorelyuk_dark_2019, redmond_implementation_2022}. The use of this algorithm allows for the DZ to be maintained despite limiting environmental factors which are discussed in Sec~\ref{subsec:1.2}. This work demonstrates the performance of DZM on the in-air High-contrast Imager for Complex Aperture Telescopes (HiCAT) testbed at the Space Telescope Science Institute (STScI)\cite{ndiaye_high-contrast_2019, soummer_high-contrast_2019, soummer_high-contrast_2022, soummer_high-contrast_2024}. The testbed configuration used in this paper consists of an apodized pupil Lyot coronagraph (APLC) with an IrisAO PTT111L DM as the segmented aperture, two Boston Micromachine (BMC) kilo DMs, and a NKT Photonics Superk EVO laser at 640~nm serving as the source. The experiments are first tested on a testbed simulator (built from the \texttt{catkit2}  package)\cite{por_spacetelescope/catkit2:_2024} and then tested on the physical testbed. The results are then compared with the theoretical bounds of HOWFSC performance based on photon noise\cite{pogorelyuk_information-theoretical_2021}.
 
\subsection{Contrast Limiting Factors}
\label{subsec:1.2}
When considering a space-based high-contrast instrument, there are a variety of variables that affect the instantaneous and steady state achievable contrast of the instrument. These variables can be separated into two categories: environmental variables and WFSC variables. This work aims to determine how a subset of these variables relate to the achievable contrast of a high-contrast instrument. 

There are a variety of environmental disturbances that impact and limit the performance of space telescopes acting on various spatial orders\cite{coyle_large_2019}. In the high spatial order regime, a mode of perturbation is DM actuator drift, arising from thermal and mechanical instabilities\cite{mejia_prada_high-contrast_2019}. Changes in the high-order regime on long time scales during observing can lead to degradation of the DZ. By injecting various drift magnitudes, we can assess the ability of the system to maintain the DZ while facing this contrast-limiting factor. 

Another important environmental limitation of DZM is the reliance on images taken in the focal plane from the science camera behind the coronagraph. For dim target stars, this could provide very low photon rates at the science camera, depending on the throughput of the coronagraph system. The Roman coronagraph images will have a photon rate of 10 milli-photons per second per pixel\cite{nemati_photon_2020}. Hence, the other important parameter that can be tested in this system is the direct photon rate at the aperture of the instrument, as prescribed from the source. Due to the Poissonian nature of the photon distribution as it arrives onto the science camera, the direct photon rate prescription is applied with Poisson noise, also known as shot noise\cite{nemati_photon_2020}. 

Focal plane sensing techniques have shown the need for the DMs to introduce phase diversity in the images to adequately measure (and then null) electric field aberrations using intensity measurements from the science camera\cite{riggs_integrated_2016}. The average amplitude of phase diversity that the DM introduces into the system is referred to in this paper as the ``dither magnitude". For a given drift value, this dither magnitude is optimized to limit contrast loss while still being able to accurately estimate electric field values from intensity measurements.

DZM has been demonstrated on the HiCAT testbed previously in Redmond et al.\ 2021\cite{redmond_dark_2021} and Redmond et al.\ 2022 \cite{redmond_dark_2022} bringing the Technology Readiness Level (TRL) of this HOWFSC algorithm to 3. Exploring the limits of performance based on the disturbances during observation while comparing these results with the theoretical bounds and quantifying differences between simulations and experiments will help raise the TRL to 4. 

An overview of DZM is described in Sec.~\ref{subsec:2.1}, the theoretical bounds process is explained in Sec.~\ref{subsec:2.2}, the process of applying the experimental parameters to DZM is presented in Sec.~\ref{subsec:2.3}, and its implementation on the HiCAT testbed is detailed in Sec.~\ref{subsec:2.4}. The results for the experiment are discussed in Sec.~\ref{sec:Results} and Sec.~\ref{sec:Conclusion} provides conclusions and outlines the future work for this project. 

\section{METHODOLOGY}
\label{sec:Methodology}

\subsection{Dark Zone Maintenance Overview}
\label{subsec:2.1}
DZM provides a closed-loop wavefront sensing and control framework to estimate and null electric field aberrations to maintain high contrast coronagraphic images\cite{pogorelyuk_dark_2019}. The mathematical formulation of the algorithm used in this paper is described in detail in Redmond et al.\ 2022\cite{redmond_implementation_2022}. Once a high contrast ``dark zone" region is dug, the DZM algorithm takes coronagraph science exposures that provide measurements to an Extended Kalman Filter (EKF). By providing a small amount of dither on the DMs to introduce phase diversity, the EKF is able to estimate the open-loop electric field using intensity measurements from a single science exposure. This estimate is used to inform the Electric Field Conjugation (EFC) controller, which provides DM commands to null these aberrations\cite{giveon_electric_2007}. As the shape of the DMs change the propagation of the wavefront, images taken by the science camera measure the change in wavefront through intensity measurements, closing the control loop and allowing the DZ to be maintained.

\begin{figure} [htb]
   \begin{center}
   \begin{tabular}{c} 
   \includegraphics[width=0.95\textwidth]{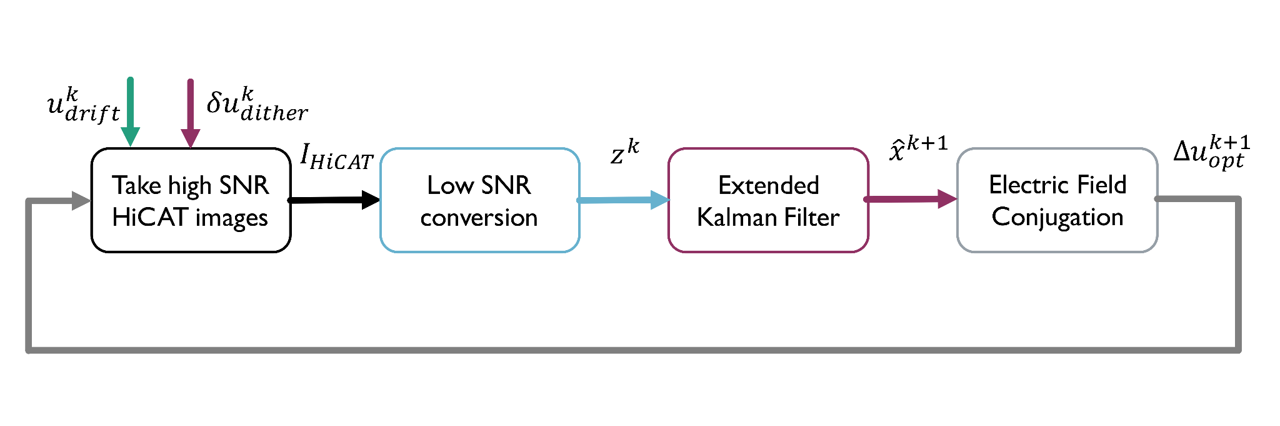}
   \end{tabular}
   \end{center}
   \caption[example] 
   {\label{fig:diagram}A diagram of the DZM algorithm control loop implemented in this experiment\cite{redmond_dark_2022}.}
\end{figure}

The DZM algorithm was adjusted in this experiment to analyze important contrast limiting factors. At each iteration, both a drift command $u_{drift}^k$ and a dither command $\delta u_{dither}^k$ are applied to the DMs which affects the measurement image taken by the science camera, $I_{HiCAT}$. This image is then degraded by a low Signal-to-Noise Ratio (SNR) conversion process to give a measurement $z^{k}$ to the EKF, which in turn provides an estimate of the electric field $\hat{x}^{k+1}$ to the EFC controller. That estimate is used to calculate an optimal command for the DM, $\Delta u_{opt}^{k+1}$, thus closing the control loop. An overview of the entire closed-loop DZM process utilized for this experiment is shown in Figure~\ref{fig:diagram}.

\newpage
\subsection{Theoretical Bounds}
\label{subsec:2.2}
Before we analyze the performance of DZM on the testbed as we vary the parameters, it is also possible to measure the theoretical bounds of performance based on the photon noise applied on the system. As explained in Pogorelyuk et al.\ 2021\cite{pogorelyuk_information-theoretical_2021}, the chosen parameters of drift and direct photon rate are used to calculate closed-form expressions of contrast. First, the $r$ closed-loop wavefront error (WFE) mode coefficients $\epsilon^{CL} \in \mathbb{R}^{r}$, are expressed as a multivariate Gaussian distribution. These modes are represented as zero-mean with a covariance expressed as 
\begin{equation}
\label{eq:epsilon}
\epsilon_{k+1}^{CL} \sim \mathcal{N}(0, P_{k} + Q) \, ,
\end{equation}
where $Q \in \mathbb{R}^{r \times r}$ is the process noise covariance matrix populated as a diagonal matrix with the drift magnitude as the standard deviation, and $P_{k} \in \mathbb{R}^{r \times r}$ is the steady-state lower bound on closed-loop covariance at every $k$ iteration. The maximum achievable contrast is expressed as 
\begin{equation}
\label{eq:Theoretical Contrast}
C =\frac{mean(\kappa\lVert G\epsilon_{k+1}^{CL}+E_{0,i} \rVert^{2})}{\kappa} \, ,
\end{equation}
where $\kappa$ is the photon rate of the DZ, calculated from the mean number of photons inside the DZ, $G$ is the Jacobian mapping the science camera pixels to the DM commands, and $E_{0,i}$ is the original static electric field. 

The theoretical contrast bounds are calculated for all possible values of mean DZ photons per pixel and drift mentioned in Sec.~\ref{subsec:2.3}. Note, the dark current is neglected due to the long exposure times chosen for the implementation.

\subsection{Parameter Implementation}
\label{subsec:2.3}
Multiple parameters impact the performance of DZM that are relevant when considering a space-based high-contrast instrument. The control loop requires generating a DM command from an electric field estimate, provided by science images affected by the direct photon rate and drift. Additionally, the dither is an internal parameter that affects the quality of the science images, which is discussed later in the paper. As previously described, a random drift command is applied at each iteration to the Boston DMs on HiCAT, providing a random walk noise profile to the system. The command is calculated via
\begin{align}
    u^k_{drift} = u^{k-1}_{drift} + \mathcal{N}(0,\sigma _{drift} ^2 \mathcal{I})
\end{align}
where $k$ is the iteration, $\sigma_{drift}$ is the standard deviation of the normal distribution $\mathcal{N}$, and $\mathcal{I}$ is the identity matrix. We characterize the drift rate using the standard deviation ($\sigma_{drift}$) and provide results for $\sigma_{drift}=0.03-3~nm/\sqrt{iteration}$. This range of values is within the range that a space-telescope could encounter, while including the upper bound conditions at which DZM would likely fail.

Additionally, for each drift condition tested, a corresponding optimal dither is found and applied to provide the phase diversity required for the EKF. The optimal dither is applied to the DMs at each iteration of DZM with a random command generated from the prescribed standard deviation $\sigma_{dither}$. Ensuring accurate measurements requires adequate phase diversity, which causes the dither values to be larger than the drift perturbations that are applied to the DMs at every iteration. Therefore, dither magnitudes of $5\times, 10\times, 15\times,$ and $20\times$ the drift magnitude are tested to characterize the electric field in the images despite aberrations, from which an optimal dither magnitude could be found for each drift case tested. A dither command is generated from the dither magnitude $\sigma_{dither}$ as

\begin{align}
    \delta u^k_{dither} = \mathcal{N}(0,\sigma _{dither} ^2 \mathcal{I})
\end{align}
for every $k$ iteration. The drift and dither magnitudes tested in the parameter scan are shown in Table~\ref{tab: Maintenace Parameters Table}.

\begin{table}[ht]
\caption{Drift rates and corresponding dither magnitudes tested in the parameter scan. Note that a maximum of 5.0~nm of dither is prescribed based on limitations of the DMs on the testbed.} 
\label{tab: Maintenace Parameters Table}
\begin{center}       
\begin{tabular}{|l|l|} 
\hline
\rule[-1ex]{0pt}{3.5ex} $\sigma_{drift}~[nm/\sqrt{iteration}]$  &  $\sigma_{dither}~[nm]$  \\
\hline
\rule[-1ex]{0pt}{3.5ex}  0.03 & 0.15, 0.3, 0.45, 0.60 \\
\hline
\rule[-1ex]{0pt}{3.5ex}  0.07 & 0.35, 0.70, 1.05, 1.40  \\
\hline
\rule[-1ex]{0pt}{3.5ex}  0.09 & 0.45, 0.90, 1.35, 1.80   \\
\hline
\rule[-1ex]{0pt}{3.5ex}  0.30 & 1.5, 3.0, 4.50, 6.0   \\
\hline
\rule[-1ex]{0pt}{3.5ex}  0.50 & 2.50, 5.0 \\
\hline
\rule[-1ex]{0pt}{3.5ex}  0.70 & 3.50, 5.0 \\
\hline
\rule[-1ex]{0pt}{3.5ex}  1.0 & 5.0  \\
\hline
\rule[-1ex]{0pt}{3.5ex}  3.0 & 5.0  \\
\hline 
\end{tabular}
\end{center}
\end{table}

\begin{figure}[htb]
    \centering
    \begin{subfigure}[b]{0.45\textwidth}
        \centering
    \includegraphics[width=0.7\textwidth]{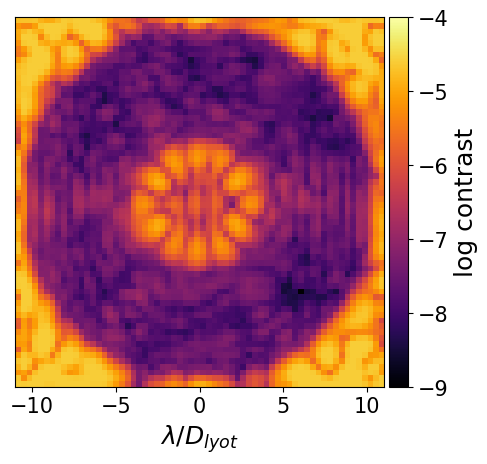}
    \end{subfigure}
    \hspace{0.05\textwidth}
    \begin{subfigure}[b]{0.45\textwidth}
        \centering
    \includegraphics[width=0.7\textwidth]{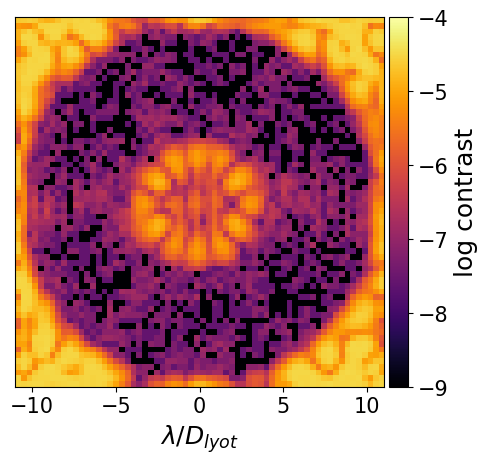}
    \end{subfigure}
    \caption{\label{fig:figure2}Left panel: Original unaltered coronagraph image on the HiCAT testbed with a DZ generated.
    Right panel: Flux-adjusted noisy HiCAT image with a direct photon rate of $1.85 \times 10^{7}$ ph/s applied.}
\end{figure}

The other parameter we consider is the direct photon rate in the system. To prescribe a direct photon rate, we follow a similar procedure to Redmond et al.\ 2022\cite{redmond_dark_2022}. A ``direct image'' (no coronagraph), without the Lyot stop, is taken on the HiCAT testbed with a very long exposure time to have the Poisson noise in the image dominate over the dark current and read noise. This image is converted from counts/s to ph/s and summed across all the pixels to find the total photon rate for a direct image on the testbed, $F_{HiCAT}$. A scaling parameter, $\gamma$, is then calculated as
\begin{align}
    \gamma = \frac{F}{F_{HiCAT}}
\end{align}
where $F$ is the prescribed direct photon rate parameter. The low photon adjustment is applied with Poisson noise on a high SNR coronagraphic image acquired by the science camera
\begin{equation}
\label{eq:Poisson}
I_{noisy} = \frac{round(P((\gamma I_{HiCAT} + \mu_{I}) t_{exposure}))}{t_{exposure}} \, ,
\end{equation}
where $I_{HiCAT}$ is the coronagraphic image converted to ph/s, $\gamma$ is the previously calculated scaling factor, $\mu_{I}$ is the dark current in the image, and $t_{exposure}$ is the exposure time chosen based on expected observing times. The dark current is chosen to be 0.005 e-/s and the exposure time is 30 s, similar to Roman observing times. A direct photon rate of $1.85 \times 10^{7}$ ph/s is chosen from the RST CGI Observing Scenario 11(47 UMa, V=5.04, G1V, $D_{star}$=0.9 mas)\cite{krist_observing_2022}. The other direct photon rates are chosen near the value of $1.85 \times 10^{7}$: $1.85 \times 10^{6}$, $8.0 \times 10^{6}$, and $1.85 \times 10^{8}$ ph/s. The real-time image-processing based on prescribed photon noise to generate images with a low SNR is shown in Figure~\ref{fig:figure2}.

When considering an instruments ability to image an exoplanet, the number of photons in the DZ is a much more relevant number than the total photon rate in a direct image. In Table~\ref{tab:Photon rate Table} we provide the initial mean number of photons per DZ pixel before dithering, $\kappa$, after the coronagraph images were flux-adjusted with a given direct photon rate. 
\FloatBarrier
\begin{table}[htb]
\caption{Initial mean number of photons per pixel per frame inside the DZ before dithering, $\kappa$, as a function of direct photon rate applied.} 
\label{tab:Photon rate Table}
\begin{center}       
\begin{tabular}{|l|l|} 
\hline
\rule[-1ex]{0pt}{3.5ex}  Direct Photon Rate Applied [ph/s]  & Initial Mean Number of DZ Photons, $\kappa$  \\
\hline
\rule[-1ex]{0pt}{3.5ex}  $1.85 \times 10^{6}$ & 0.38 \\
\hline
\rule[-1ex]{0pt}{3.5ex}  $8.0 \times 10^{6}$ & 0.47 \\
\hline
\rule[-1ex]{0pt}{3.5ex}  $1.85 \times 10^{7}$ & 0.59 \\
\hline
\rule[-1ex]{0pt}{3.5ex}  $1.85 \times 10^{8}$ & 1.41 \\
\hline
\end{tabular}
\end{center}
\end{table}

\subsection{HiCAT Testbed Implementation}
\label{subsec:2.4}
The HiCAT testbed located at STScI is used to run DZM experiments while varying different system parameters. A DZ is first dug using pair-wise probing (PWP) as the estimation technique which introduces phase diversity into the science exposures to measure electric field aberrations \cite{shaklan_pair-wise_2011}. The EFC controller is fed the estimate to null the static aberrations and dig a DZ in a recursive manner to achieve a high-quality contrast. After 250 iterations, the mean DZ contrast achieved by PWP is $3\times10^{-9}$ in simulator and $3\times10^{-8}$ in hardware, as the HiCAT simulator is yet to be optimized to model testbed imperfections. The PWP+EFC algorithm is then terminated and the DZ is maintained for a given combination of drift, dither, and direct photon rate over a duration of 1000 iterations. The Jacobian for the testbed is previously characterized to map the pixel electric field measurements from the science camera to the DM commands. For all DZM experiments, a long coronagraphic exposure time is required to minimize effects of the detector noise relative to the incoming photon noise and provide an image with good SNR. However, due to the large dither values applied, the leaking of light in poor contrast conditions to the DZ could lead to pixel saturation of the science camera. An acceptable exposure time can be calculated as
\begin{equation}
\label{eq:Exposure Time}
t_{exp} = \frac{p (2^{16}-1)}{f_{max}} \, ,
\end{equation}
where $p$ is the well depth percentage target for the science camera of 65\% and $f_{max}$ is the maximum flux in the DZ for a given dither value. While Eq.~\ref{eq:Exposure Time} was applied for larger dither values, it leads to very large exposure times for smaller dithers, on the order of magnitude of a few seconds. To limit the duration of the experiment, a maximum exposure time of 100,000 microseconds was chosen. For the direct images, an auto-exposure function was applied to avoid saturating the pixels from the bright source on the HiCAT testbed. Before running the experiments on the testbed, the HiCAT simulator was used to test and validate functionality and get an initial estimate of performance for DZM across different system parameters.
The dimensions of the DZ along with the binning factor applied to the raw images and the coronagraph exposure times are shown in Table~\ref{tab:Testbed Parameters Table}. 
\begin{table}[ht]
\caption{HiCAT Testbed configuration for PWP and DZM experiments.} 
\label{tab:Testbed Parameters Table}
\begin{center}       
\begin{tabular}{|l|l|l|} 
\hline
\rule[-1ex]{0pt}{3.5ex} Parameter & PWP  & DZM  \\
\hline
\rule[-1ex]{0pt}{3.5ex}  DZ Inner-Working Angle (iwa) & 3.7 & 4 \\
\hline
\rule[-1ex]{0pt}{3.5ex}  DZ Outer-Working Angle (owa) & 10 & 9 \\
\hline
\rule[-1ex]{0pt}{3.5ex}  Science Exposure Binning Factor & 3 & 3 \\
\hline
\rule[-1ex]{0pt}{3.5ex}  Coronagraph Exposure Time [$\mu$s] & 5,000 & 100,000 \\
\hline
\end{tabular}
\end{center}
\end{table}

\section{RESULTS}
\label{sec:Results}
\subsection{Theory}
\begin{figure} [ht]
   \begin{center}
   \begin{tabular}{c} 
   \includegraphics[height=5cm]{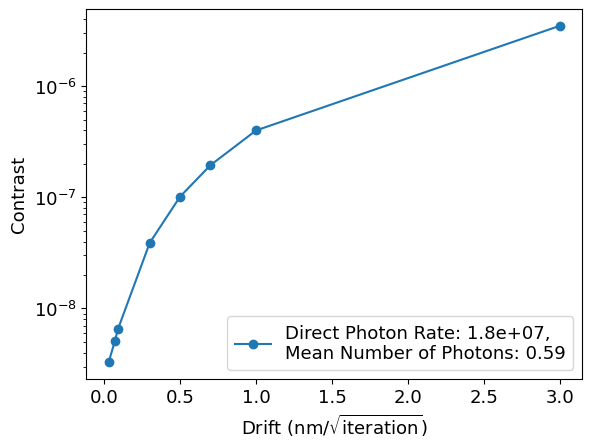}
   \end{tabular}
   \end{center}
   \caption[example] 
   {\label{fig:theory} 
Theoretical photon noise bounds for a 
$\kappa$ of 0.59. }
\end{figure}

In this section, the results comparing the performance of DZM with varying system parameters are shown using photon-noise theoretical bounds, the data from the HiCAT simulator, and the data collected from the HiCAT testbed. Figure~\ref{fig:theory} illustrates the theoretical photon noise bounds of possible contrast achievable in the DZ as a function of only the drift and the mean number of photons in the DZ $\kappa$, which is computed from the coronagraph images with the direct photon rate prescribed. 

\subsection{Simulation}
\begin{figure}[ht]
    \centering
    \begin{subfigure}[b]{0.45\textwidth}
        \centering
    \includegraphics[width=\textwidth]{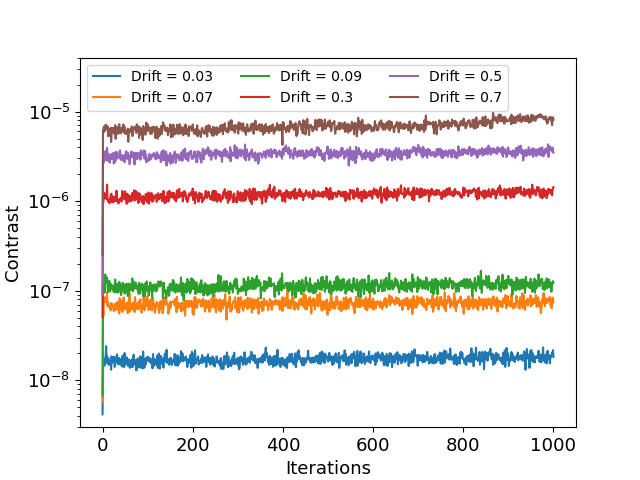}
    \end{subfigure}
    \hspace{0.05\textwidth}
    \begin{subfigure}[b]{0.45\textwidth}
        \centering
    \includegraphics[width=\textwidth]{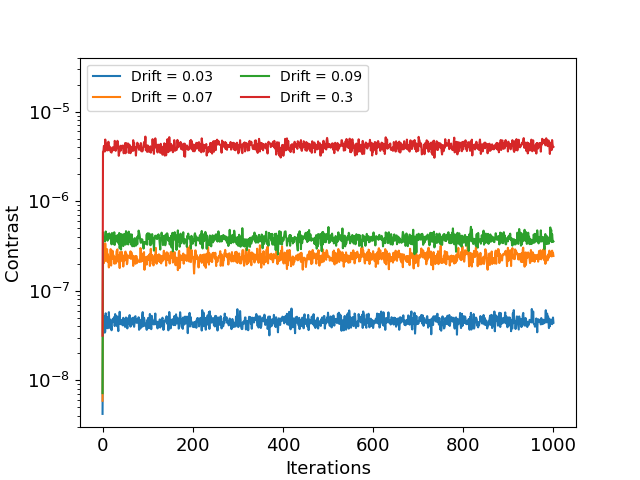}
    \end{subfigure}
    \caption{\label{fig:simulation} Left panel: Simulation contrast runs with a $\kappa$ of 0.59 and a dither factor of $5\times$ the drift rate. Right panel: Simulation contrast runs with a $\kappa$ of 0.59 and a dither factor of $10\times$ the drift rate.}
\end{figure}

The simulation results are then run to verify functionality of the parameter scan before experiments are run on the testbed, along with providing a benchmark for possible performance. After digging a DZ using PWP, the parameter scan is run for all combinations of drift, dither, and direct photon rate. The performance of DZM is analyzed and is illustrated for the case where the $\kappa$ was 0.59 in Figure~\ref{fig:simulation}, displaying the contrast as a function of drift for a data subset with the cases of dither being 5$\times$ and 10$\times$ the drift. The experiments with the dither being 5$\times$ the drift provide better performance in simulation for every combination of drift and direct photon rate. 

\subsection{Testbed}
\begin{figure}[ht]
    \centering
    \begin{subfigure}[b]{0.45\textwidth}
        \centering
    \includegraphics[width=\textwidth]{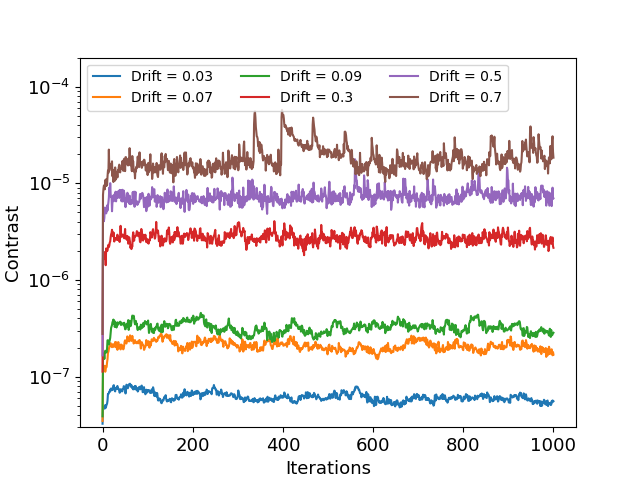}
    \end{subfigure}
    \hspace{0.05\textwidth}
    \begin{subfigure}[b]{0.45\textwidth}
        \centering
    \includegraphics[width=\textwidth]{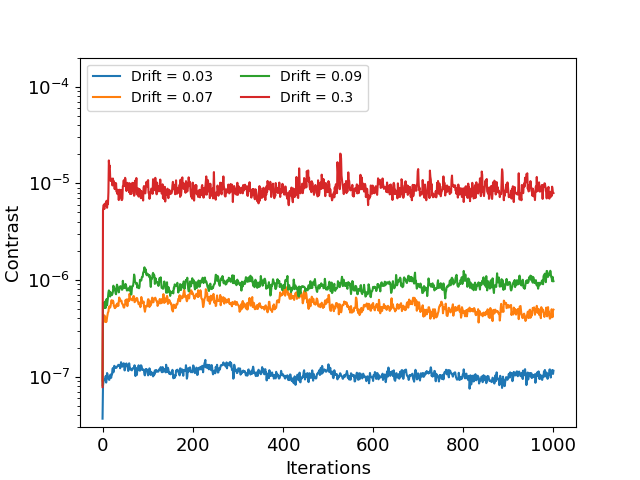}
    \end{subfigure}
    \caption{\label{fig:hw} Left panel: Hardware contrast runs for a $\kappa$ value of 0.59 and a dither factor of $5\times$ the drift rate. Right panel: Hardware contrast runs for a $\kappa$ value of 0.59 and a dither factor of $10\times$ the drift rate.}
\end{figure}

With references from the theoretical bounds and the simulation runs, the parameter scan was then run on the HiCAT testbed. The pair-wise probing allowed all DZM runs to begin with a contrast of $3 \times 10^{-8}$, and the results achieved are shown in Figure~\ref{fig:hw} for a subset of cases with a $\kappa$ of 0.59 with the dither being $5\times$ and $10\times$ the drift.

The results demonstrate that the simulation was able to maintain a more stable contrast level than the hardware, most likely due to the presence of small air turbulence in the testbed along with incoherent light, both features not modeled in simulation. For both simulation and hardware, the optimal dither is identified as the case providing the lowest and most consistent mean contrast for a given drift and direct photon rate combination. In hardware the optimal dither was 5$\times$ the drift for all cases except for two in which it was 10$\times$: when the drift was 0.03 nm and the direct photon rate was $1.85\times10^{6}$ ph/s and when the drift was 0.7 nm and the direct photon rate was $1.85\times10^{8}$ ph/s.

\subsection{Comparison}
\begin{figure}[htb]
    \centering
    \begin{subfigure}[b]{0.45\textwidth}
        \centering
    \includegraphics[width=\textwidth]{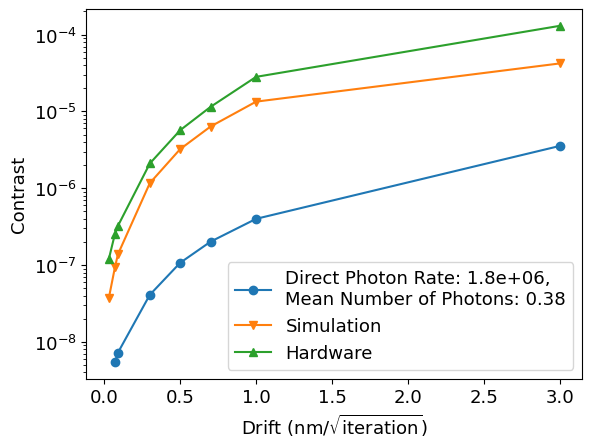}
    \end{subfigure}
    \hspace{0.05\textwidth}
    \begin{subfigure}[b]{0.45\textwidth}
        \centering
    \includegraphics[width=\textwidth]{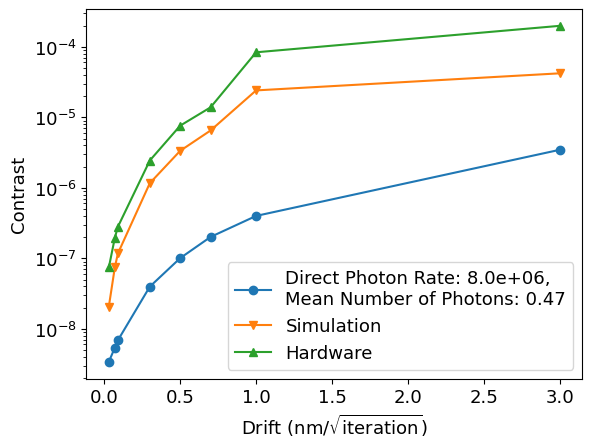}
    \end{subfigure}
    \centering
    \vspace{0.05\textwidth}
    \begin{subfigure}[b]{0.45\textwidth}
        \centering
    \includegraphics[width=\textwidth]{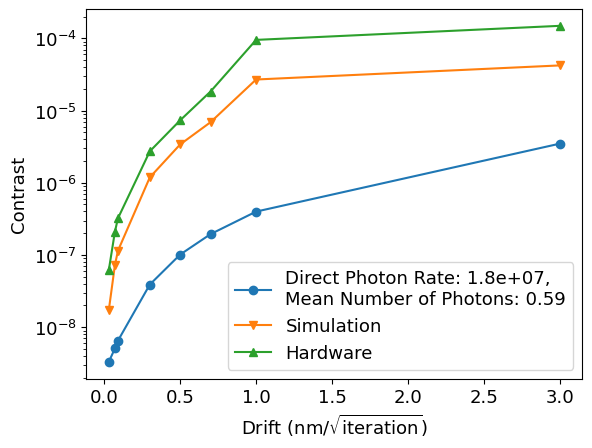}
    \end{subfigure}
    \hspace{0.05\textwidth}
    \centering
    \begin{subfigure}[b]{0.45\textwidth}
        \centering
    \includegraphics[width=\textwidth]{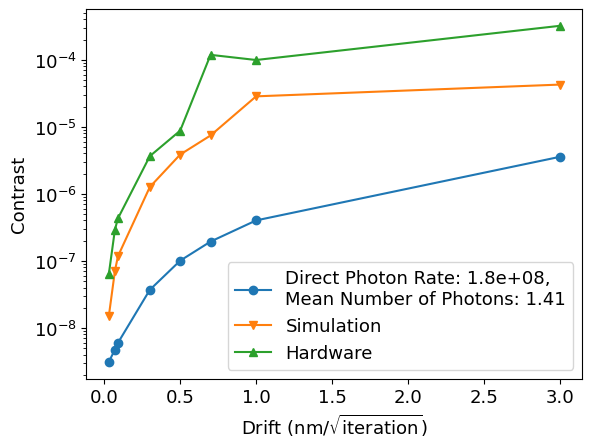}
    \end{subfigure}
    \caption{\label{fig:Combinede6} Mean contrast performance of theoretical bounds, hardware, and software runs at optimal dither for a direct photon rate of - Top left: $1.85 \times 10^{6}$~ph/s and a $\kappa$ of 0.38. Top right: $8.0 \times 10^{6}$ ph/s and a $\kappa$ of 0.47. Bottom left: $1.85 \times 10^{7}$ ph/s and a $\kappa$ of 0.59. Bottom right: $1.85 \times 10^{8}$ ph/s and a $\kappa$ of 1.41.}

\end{figure}

Comparing the performance of hardware, software, and theory there is almost a one and a half order of magnitude difference between the theoretical photon noise bounds and the performance of simulation and hardware, indicating the opportunity for advanced algorithms to close this gap. This discrepancy also grows non-linearly with the drift, showing possible limitations of the EKF in high drift regimes. Another important disparity is the contrast difference between simulation and hardware, especially near the lower drift regime, with this gap closing as drift increases. There are also small differences in achieved contrast levels and the consistency of contrast maintained that need to be further investigated. When comparing all direct photon rates, it is evident that the contrast achieved by DZM is affected more by the drift magnitude applied to the DMs than the magnitude of the direct photon rate and the subsequent noisy images passed on to the EKF as measurements. These results are illustrated in Figure~\ref{fig:Combinede6}. 

A summary of the experiments, providing a direct comparison between the three sources of data, with the simulation and hardware data showing the mean contrast with standard deviation, along with the mean number of photons in the DZ for the hardware runs, is shown in Table~\ref{tab:Results Table}. The contrast of the dither case providing the best performance is reported in the table. 

\clearpage
\begin{table}[H]
\caption{Summary of the results showing the theoretical contrast bounds along with the simulation and hardware mean contrast achieved with its standard deviation. The dither giving the best contrast performance is chosen. For the hardware runs, the temporally averaged mean number of photons per pixel inside the DZ is also reported.} 
\label{tab:Results Table}
\begin{center}
\resizebox{\textwidth}{!}{%
\begin{tabular}{|l|l|l|l|l|l|} 
\hline
\rule[-1ex]{0pt}{3.5ex} Drift ($nm/\sqrt{iteration}$)            & Direct Photon Rate (ph/s)   & Theoretical Bound &  Simulation Contrast & Hardware Contrast & Mean Photon \\
\hline
\rule[-1ex]{0pt}{3.5ex} \multirow{6}{*}{0.03} & $1.85 \times 10^{6}$ &  $3.49 \times 10^{-9}$ & $3.72 \times 10^{-8}\pm 6.18\times 10^{-9}$ & $1.23 \times 10^{-7}\pm 1.14\times 10^{-8}$ & 0.82\\
                                              & $8.0 \times 10^{6}$    & $3.40 \times 10^{-9}$ & $2.07 \times 10^{-8}\pm 2.23\times 10^{-9}$ & $7.41 \times 10^{-8}\pm 8.79\times 10^{-9}$ & 1.22\\
                                              & $1.85 \times 10^{7}$ & $3.31 \times 10^{-9}$ & $1.73 \times 10^{-8}\pm 1.87\times 10^{-9}$ & $6.20 \times 10^{-8}\pm 6.91\times 10^{-9}$ & 1.69\\
                                              & $1.85 \times 10^{8}$ & $3.12 \times 10^{-9}$ & $1.50 \times 10^{-8}\pm 1.60\times 10^{-9}$ & $6.34 \times 10^{-8}\pm 1.05\times 10^{-8}$ & 5.16\\
\hline
\rule[-1ex]{0pt}{3.5ex} \multirow{6}{*}{0.07} & $1.85 \times 10^{6}$ &  $5.50 \times 10^{-9}$ & $9.31 \times 10^{-8}\pm 1.08\times 10^{-8}$ & $2.54 \times 10^{-7}\pm 3.78\times 10^{-8}$ & 1.09\\
                                              & $8.0 \times 10^{6}$    & $5.43 \times 10^{-9}$ &  $7.54 \times 10^{-8}\pm 8.53\times 10^{-9}$ & $1.97\times 10^{-7}\pm 2.92\times 10^{-8}$ & 1.92\\
                                              & $1.85 \times 10^{7}$ & $5.07 \times 10^{-9}$ &  $7.22 \times 10^{-8}\pm 7.96\times 10^{-9}$ & $2.09 \times 10^{-7}\pm 2.51\times 10^{-8}$ & 2.93\\
                                              & $1.85 \times 10^{8}$ & $4.73 \times 10^{-9}$ & $7.04 \times 10^{-8}\pm 8.15\times 10^{-9}$ & $2.91 \times 10^{-7}\pm 5.23\times 10^{-8}$ & 10.9\\
\hline
\rule[-1ex]{0pt}{3.5ex} \multirow{6}{*}{0.09} & $1.85 \times 10^{6}$ & $7.24 \times 10^{-9}$ & $1.38 \times 10^{-7}\pm 1.59\times 10^{-8}$ & $3.24 \times 10^{-7}\pm 3.86\times 10^{-8}$ & 1.22\\
                                              & $8.0 \times 10^{6}$    & $6.89 \times 10^{-9}$ & $1.18 \times 10^{-7}\pm 1.42\times 10^{-8}$ & $2.81 \times 10^{-7}\pm 4.96\times 10^{-8}$ & 2.22\\
                                              & $1.85 \times 10^{7}$ & $6.49 \times 10^{-9}$ & $1.14 \times 10^{-7}\pm 1.36\times 10^{-8}$ & $3.21 \times 10^{-7}\pm 4.39\times 10^{-8}$ & 3.51\\
                                              & $1.85 \times 10^{8}$ & $5.97 \times 10^{-9}$ & $1.17 \times 10^{-7}\pm 1.45\times 10^{-8}$ & $4.29 \times 10^{-7}\pm 8.52\times 10^{-8}$ & 12.81\\
\hline
\rule[-1ex]{0pt}{3.5ex} \multirow{6}{*}{0.3} & $1.85 \times 10^{6}$  & $4.10 \times 10^{-8}$ & $1.17 \times 10^{-6}\pm 1.14\times 10^{-7}$ & $2.10 \times 10^{-6}\pm 2.33\times 10^{-7}$ & 2.98\\
                                              & $8.0 \times 10^{6}$    & $3.97 \times 10^{-8}$ & $1.16 \times 10^{-6}\pm 1.20\times 10^{-7}$ & $2.44 \times 10^{-6}\pm 3.16\times 10^{-7}$ & 6.49\\
                                              & $1.85 \times 10^{7}$ & $3.90 \times 10^{-8}$ & $1.19 \times 10^{-6}\pm 1.15\times 10^{-7}$ & $2.73 \times 10^{-6}\pm 3.55\times 10^{-7}$ & 10.43\\
                                              & $1.85 \times 10^{8}$ & $3.69 \times 10^{-8}$ &  $1.26 \times 10^{-6}\pm 1.37\times 10^{-7}$ & $3.64 \times 10^{-6}\pm 2.57\times 10^{-6}$ & 45.83\\
\hline
\rule[-1ex]{0pt}{3.5ex} \multirow{6}{*}{0.5} & $1.85 \times 10^{6}$  & $1.07 \times 10^{-7}$ & $3.23 \times 10^{-6}\pm 3.14\times 10^{-7}$ & $5.70 \times 10^{-6}\pm 6.42\times 10^{-7}$ & 4.83\\
                                              & $8.0 \times 10^{6}$    & $1.01 \times 10^{-7}$ &  $3.35 \times 10^{-6}\pm 3.28\times 10^{-7}$ & $7.60 \times 10^{-6}\pm 1.15\times 10^{-6}$ & 11.46\\
                                              & $1.85 \times 10^{7}$ & $1.01 \times 10^{-7}$ & $3.40 \times 10^{-6}\pm 3.26\times 10^{-7}$ & $7.33 \times 10^{-6}\pm 1.25\times 10^{-6}$ & 16.68\\
                                              & $1.85 \times 10^{8}$ & $1.01 \times 10^{-7}$ & $3.85 \times 10^{-6}\pm 6.07\times 10^{-7}$ & $8.75 \times 10^{-6}\pm 2.36\times 10^{-6}$ & 56.05\\
\hline
\rule[-1ex]{0pt}{3.5ex} \multirow{6}{*}{0.7} & $1.85 \times 10^{6}$  & $2.02 \times 10^{-7}$ & $6.36 \times 10^{-6}\pm 6.63\times 10^{-7}$ & $1.15 \times 10^{-5}\pm 1.53\times 10^{-6}$ & 6.82\\
                                              & $8.0 \times 10^{6}$    & $2.01 \times 10^{-7}$ & $6.50 \times 10^{-6}\pm 6.84\times 10^{-7}$ & $1.38 \times 10^{-5}\pm 3.23\times 10^{-6}$ & 15.58\\
                                              & $1.85 \times 10^{7}$ & $1.94 \times 10^{-7}$ & $6.90 \times 10^{-6}\pm 8.94\times 10^{-7}$ & $1.83 \times 10^{-5}\pm 6.32\times 10^{-6}$ & 27.2\\
                                              & $1.85 \times 10^{8}$ & $1.94 \times 10^{-7}$ & $7.41 \times 10^{-6}\pm 8.79\times 10^{-7}$ & $1.18 \times 10^{-4}\pm 9.21\times 10^{-5}$ & 110.42\\
\hline
\rule[-1ex]{0pt}{3.5ex} \multirow{6}{*}{1.0} & $1.85 \times 10^{6}$  & $4.00 \times 10^{-7}$ & $1.34 \times 10^{-5}\pm 1.30\times 10^{-6}$ & $2.82 \times 10^{-5}\pm 5.12\times 10^{-6}$ & 10.27\\
                                              & $8.0 \times 10^{6}$    & $3.99 \times 10^{-7}$ & $2.41 \times 10^{-5}\pm 2.08\times 10^{-6}$ & $8.35 \times 10^{-5}\pm 4.69\times 10^{-5}$ & 36.37\\
                                              & $1.85 \times 10^{7}$ & $3.99 \times 10^{-7}$ & $2.69 \times 10^{-5}\pm 2.35\times 10^{-6}$ & $9.51 \times 10^{-5}\pm 6.04\times 10^{-5}$ & 57.43\\
                                              & $1.85 \times 10^{8}$ & $4.01 \times 10^{-7}$ & $2.84 \times 10^{-5}\pm 2.98\times 10^{-6}$ & $9.90 \times 10^{-5}\pm 6.55\times 10^{-5}$ & 192.32\\
\hline
\rule[-1ex]{0pt}{3.5ex} \multirow{6}{*}{3.0} & $1.85 \times 10^{6}$  & $3.55 \times 10^{-6}$ & $4.21 \times 10^{-5}\pm 4.60\times 10^{-6}$ & $1.30 \times 10^{-4}\pm 5.28\times 10^{-5}$ & 22.47\\
                                              & $8.0 \times 10^{6}$    &  $3.46 \times 10^{-6}$ & $4.20 \times 10^{-5}\pm 4.94\times 10^{-6}$ & $1.98 \times 10^{-4}\pm 7.51\times 10^{-5}$ & 57.87\\
                                              & $1.85 \times 10^{7}$ & $3.48 \times 10^{-6}$ & $4.21 \times 10^{-5}\pm 4.15\times 10^{-6}$ & $1.49 \times 10^{-4}\pm 6.54\times 10^{-5}$ & 71.38\\
                                              & $1.85 \times 10^{8}$ & $3.56 \times 10^{-6}$ &  $4.25 \times 10^{-5}\pm 4.58\times 10^{-6}$ & $3.20 \times 10^{-4}\pm 1.96\times 10^{-4}$ & 336.17\\
\hline
\end{tabular}
}
\end{center}
\end{table}

\section{CONCLUSIONS AND FUTURE WORK}
\label{sec:Conclusion}
This work demonstrates the performance of DZM under varying system conditions, namely the drift magnitude of the DMs and the direct photon rate entering the primary mirror, driving the shot noise of the measurements passed onto the EKF. The effects of these parameters were tested through multiple platforms, including the theoretical photon bounds, the HiCAT simulator (\texttt{catkit2}\cite{por_spacetelescope/catkit2:_2024}), and the HiCAT testbed at STScI. Spanning a large range of drift rates from $0.03-3~nm/\sqrt{iteration}$ along with a direct photon rate from $1.85\times10^{6}$--$1.85\times10^{8}$ ph/s, there is a mean difference between the theoretical photon bounds and the performance on the testbed of approximately one and a half orders of magnitude. 

Future work will aim to reduce this difference through the use of advanced algorithms, namely a modal EKF as explained in Pogorelyuk et al.\ 2022\cite{pogorelyuk_dark_2022}, that will model and estimate the modes of the electric fields, rather than a single pixel EKF, taking into account pixel cross-talk. To ensure the system is properly characterized, Expectation-Maximization will also be used to perform system identification and update the Jacobian during maintenance as described in Sun et al.\ 2018\cite{sun_identification_2018}, reducing the discrepancy between simulation and hardware. Finally, while this maintenance parameter scan has been run on the HiCAT air testbed, it along with the above mentioned advanced algorithms will be run on the vacuum Decadal Survey Testbed (DST) at JPL in an attempt to demonstrate $10^{-10}$ performance, thus increasing the TRL of DZM to 4. This will also allow for a comparison of performance between DZM and slewing.
%\appendix    %>>>> this command starts appendixes

%\section{Further plots?}
%\label{sec:misc}

\acknowledgments % equivalent to \section*{ACKNOWLEDGMENTS}  
 
This work was supported by the National Aeronautics and Space Administration under Grant NNH21ZDA001N issued through the Strategic Astrophysics Technology/Technology Demonstration for Exo-planet Missions Program (SAT-TDEM; PI: K. Cahoy). 

The HiCAT testbed has been developed over the past 10 years and benefited from the work of an extended collaboration of over 50 people. This work was supported in part by the National Aeronautics and Space Administration under Grant 80NSSC19K0120 issued through the Strategic Astrophysics Technology/Technology Demonstration for Exo-planet Missions Program (SAT-TDEM; PI: R. Soummer), and under Grant 80NSSC22K0372 issued through the Astrophysics Research and Analysis Program (APRA; PI: L. Pueyo).

E.H.P. was supported in part by the NASA Hubble Fellowship grant HST-HF2-51467.001-A awarded by the Space Telescope Science Institute, which is operated by the Association of Universities for Research in Astronomy, incorporated, under NASA contract NAS5-26555. Iva Laginja acknowledges partial support from a postdoctoral fellowship issued by the Centre National d’Etudes Spatiales (CNES) in France.

% References
\bibliography{citations} % bibliography data in report.bib

\begin{thebibliography}{10}

\bibitem{stark_exoearth_2019}
Stark, C.~C., ``{ExoEarth} yield landscape for future direct imaging space telescopes,'' {\em Journal of Astronomical Telescopes, Instruments, and Systems}~{\bf 5},  1 (May 2019).

\bibitem{seo_testbed_2019}
Seo, B.-J., Patterson, K., Balasubramanian, K., Crill, B., Chui, T., Echeverri, D., Kern, B.~D., Marx, D., Moody, D., Mejia~Prada, C., Ruane, G., Shi, F., Shaw, J., Siegler, N., Tang, H., Trauger, J., Wilson, D.~W., and Zimmer, R., ``Testbed demonstration of high-contrast coronagraph imaging in search for {Earth}-like exoplanets,'' in [{\em Techniques and {Instrumentation} for {Detection} of {Exoplanets} {IX}}{\nolinebreak\hspace{0.1em}]},  Shaklan, S.~B., ed.,  53, SPIE, San Diego, United States (Sept. 2019).

\bibitem{kasdin_nancy_2020}
Kasdin, N.~J., Bailey, V., Mennesson, B., Zellem, R., Ygouf, M., Rhodes, J., Luchik, T., Zhao, F., Riggs, A. J.~E., Seo, B.-J., Krist, J., Kern, B.~D., Tang, H., Nemati, B., Groff, T.~D., Zimmerman, N.~T., Macintosh, B.~A., Turnbull, M., Debes, J., Douglas, E.~S., and Lupu, R.~E., ``The {Nancy} {Grace} {Roman} {Space} {Telescope} {Coronagraph} {Instrument} ({CGI}) technology demonstration,'' in [{\em Space {Telescopes} and {Instrumentation} 2020: {Optical}, {Infrared}, and {Millimeter} {Wave}}{\nolinebreak\hspace{0.1em}]},  Lystrup, M., Batalha, N., Tong, E.~C., Siegler, N., and Perrin, M.~D., eds.,  194, SPIE, Online Only, United States (Dec. 2020).

\bibitem{meeker_twin_2021}
Meeker, S.~R., Noyes, M., Tang, H., Ruane, G., Prada, C.~M., Bendek, E., Baxter, W., Crill, B., Riggs, A. J.~E., Poon, P.~K., and Siegler, N., ``The {Twin} decadal survey testbeds in the high contrast imaging testbed facility at {NASA}’s jet propulsion laboratory,'' in [{\em Techniques and {Instrumentation} for {Detection} of {Exoplanets} {X}}{\nolinebreak\hspace{0.1em}]},   {\bf 11823},  322--334, SPIE (Sept. 2021).

\bibitem{pogorelyuk_dark_2019}
Pogorelyuk, L. and Kasdin, N.~J., ``Dark {Hole} {Maintenance} and {A} {Posteriori} {Intensity} {Estimation} in the {Presence} of {Speckle} {Drift} in a {High}-{Contrast} {Space} {Coronagraph},'' (2019).

\bibitem{redmond_implementation_2022}
Redmond, S.~F., Pogorelyuk, L., Pueyo, L., Por, E., Noss, J., Will, S.~D., Laginja, I., Brooks, K., Maclay, M., Fowler, J., Jeremy~Kasdin, N., Perrin, M.~D., and Soummer, R., ``Implementation of a dark zone maintenance algorithm for speckle drift correction in a high contrast space coronagraph,'' {\em Journal of Astronomical Telescopes, Instruments, and Systems}~{\bf 8} (Sept. 2022).

\bibitem{ndiaye_high-contrast_2019}
N'Diaye, M., Mazoyer, J., Choquet, E., Pueyo, L., Perrin, M.~D., Egron, S., Leboulleux, L., Levecq, O., Carlotti, A., Long, C.~A., Lajoie, R., and Soummer, R., ``High-contrast imager for {Complex} {Aperture} {Telescopes} ({HiCAT}): 3. first lab results with wavefront control,'' (2019).

\bibitem{soummer_high-contrast_2019}
Soummer, R., Brady, G.~R., Brooks, K., Comeau, T., Choquet, E., Dillon, T., Egron, S., Gontrum, R., Hagopian, J., Laginja, I., Leboulleux, L., Perrin, M.~D., Petrone, P., Pueyo, L., Mazoyer, J., N'Diaye, M., Riggs, A. J.~E., Shiri, R., Sivaramakrishnan, A., Laurent, K.~S., Valenzuela, A.-M., and Zimmerman, N.~T., ``High-contrast imager for complex aperture telescopes ({HiCAT}): 5. first results with segmented-aperture coronagraph and wavefront control,'' (2019).

\bibitem{soummer_high-contrast_2022}
Soummer, R., Por, E.~H., Pourcelot, R., Redmond, S.~F., Laginja, I., Will, S.~D., Perrin, M.~D., Pueyo, L., Sahoo, A., Petrone, P., Brooks, K.~J., Fox, R., Klein, A., Nickson, B., Comeau, T., Ferrari, M., Gontrum, R., Hagopian, J., Leboulleux, L., Leongomez, D., Lugten, J., Mugnier, L.~M., N'Diaye, M., Nguyen, M., Noss, J., Sauvage, J.-F., Scott, N., Sivaramakrishnan, A., Subedi, H.~B., and Weinstock, S., ``High-contrast imager for complex aperture telescopes ({HiCAT}): 8. {Dark} zone demonstration with simultaneous closed loop low-order wavefront sensing and control,'' in [{\em Space {Telescopes} and {Instrumentation} 2022: {Optical}, {Infrared}, and {Millimeter} {Wave}}{\nolinebreak\hspace{0.1em}]},  Coyle, L.~E., Perrin, M.~D., and Matsuura, S., eds.,  81, SPIE, Montréal, Canada (Aug. 2022).

\bibitem{soummer_high-contrast_2024}
Soummer, R., Pourcelot, R., Por, E., Steiger, S., Laginja, I., Buralli, B., Pueyo, L., Nguyen, M., Nickson, B., Sahoo, A., and {the extended HiCAT team}, ``High-contrast imager for complex aperture telescopes ({HiCAT}): 11. {System}-level static and dynamic demonstration of the {Apodized} {Pupil} {Lyot} {Coronagraph} with a segmented aperture.,'' {\em {\textbackslash}procspie} {\bf 13092} (2024).

\bibitem{por_spacetelescope/catkit2:_2024}
``spacetelescope/catkit2: v0.6.1,'' (May 2024).

\bibitem{pogorelyuk_information-theoretical_2021}
Pogorelyuk, L., Pueyo, L., Males, J.~R., Cahoy, K., and Kasdin, N.~J., ``Information-theoretical {Limits} of {Recursive} {Estimation} and {Closed}-loop {Control} in {High}-contrast {Imaging},'' {\em The Astrophysical Journal Supplement Series}~{\bf 256},  39 (Oct. 2021).

\bibitem{coyle_large_2019}
Coyle, L.~E., Knight, J.~S., Pueyo, L., Arenberg, J.~W., Bluth, A.~M., East, M., Patton, K., and Bolcar, M.~R., ``Large ultra-stable telescope system study,'' in [{\em {UV}/{Optical}/{IR} {Space} {Telescopes} and {Instruments}: {Innovative} {Technologies} and {Concepts} {IX}}{\nolinebreak\hspace{0.1em}]},  Breckinridge, J.~B., Stahl, H.~P., and Barto, A.~A., eds.,  27, SPIE, San Diego, United States (Sept. 2019).

\bibitem{mejia_prada_high-contrast_2019}
Mejia~Prada, C., Serabyn, E., and Shi, F., ``High-contrast imaging stability using {MEMS} deformable mirror,'' in [{\em Techniques and {Instrumentation} for {Detection} of {Exoplanets} {IX}}{\nolinebreak\hspace{0.1em}]},  Shaklan, S.~B., ed.,  10, SPIE, San Diego, United States (Sept. 2019).

\bibitem{nemati_photon_2020}
Nemati, B., ``Photon counting and precision photometry for the {Roman} {Space} {Telescope} {Coronagraph},'' in [{\em Space {Telescopes} and {Instrumentation} 2020: {Optical}, {Infrared}, and {Millimeter} {Wave}}{\nolinebreak\hspace{0.1em}]},  Lystrup, M., Batalha, N., Tong, E.~C., Siegler, N., and Perrin, M.~D., eds.,  271, SPIE, Online Only, United States (Dec. 2020).

\bibitem{riggs_integrated_2016}
Riggs, A. J.~E., ``Integrated {Wavefront} {Correction} and {Bias} {Estimation} for the {High}-{Contrast} {Imaging} of {Exoplanets},'' (2016).

\bibitem{redmond_dark_2021}
Redmond, S.~M., Pueyo, L., Pogorelyuk, L., Por, E.~H., Noss, J., Laginja, I., Brooks, K.~J., Perrin, M.~D., Soummer, R., and Kasdin, J., ``Dark zone maintenance results for segmented aperture wavefront error drift in a high contrast space coronagraph,'' in [{\em Techniques and {Instrumentation} for {Detection} of {Exoplanets} {X}}{\nolinebreak\hspace{0.1em}]},  Shaklan, S.~B. and Ruane, G.~J., eds.,  48, SPIE, San Diego, United States (Sept. 2021).

\bibitem{redmond_dark_2022}
Redmond, S.~F., Pueyo, L., Pogorelyuk, L., Por, E.~H., Noss, J., Brooks, K.~J., Laginja, I., Perrin, M.~D., Soummer, R., and Kasdin, J.~N., ``Dark zone maintenance for future coronagraphic space missions,'' in [{\em Space {Telescopes} and {Instrumentation} 2022: {Optical}, {Infrared}, and {Millimeter} {Wave}}{\nolinebreak\hspace{0.1em}]},  Coyle, L.~E., Perrin, M.~D., and Matsuura, S., eds.,  86, SPIE, Montréal, Canada (Aug. 2022).

\bibitem{giveon_electric_2007}
Give'on, A., Kern, B., Shaklan, S., Moody, D.~C., and Pueyo, L., ``Electric {Field} {Conjugation} - {A} {Broadband} {Wavefront} {Correction} {Algorithm} {For} {High}-contrast {Imaging} {Systems},'' ~{\bf 211},  135.20 (Dec. 2007).
\newblock ADS Bibcode: 2007AAS...21113520G.

\bibitem{krist_observing_2022}
Krist, J., ``Observing {Scenario} ({OS}) 11 time series simulations for the {Hybrid} {Lyot} {Coronagraph} {Band} 1,'' (Feb. 2022).

\bibitem{shaklan_pair-wise_2011}
Give'on, A., Kern, B.~D., and Shaklan, S., ``Pair-wise, deformable mirror, image plane-based diversity electric field estimation for high contrast coronagraphy,''  815110 (Sept. 2011).

\bibitem{pogorelyuk_dark_2022}
Pogorelyuk, L., Krist, J., Nemati, B., Riggs, A. J.~E., Miller, S., Pueyo, L., Kasdin, N.~J., and Cahoy, K., ``Dark hole maintenance with modal pairwise probing in numerical simulations of {Roman} coronagraph instrument,'' {\em Journal of Astronomical Telescopes, Instruments, and Systems}~{\bf 8} (Mar. 2022).

\bibitem{sun_identification_2018}
Sun, H., Kasdin, N.~J., and Vanderbei, R., ``Identification and adaptive control of a high-contrast focal plane wavefront correction system,'' {\em Journal of Astronomical Telescopes, Instruments, and Systems}~{\bf 4},  1 (Dec. 2018).

\end{thebibliography}
\bibliographystyle{spiebib} % makes bibtex use spiebib.bst

\end{document}